Gyrosonics a Novel Stimulant for Autonomic Nervous System

S. K. Ghatak[1], S.S.Ray[2] ,R.Choudhuri [3]& S.Banerjee[3]

[1]*Department of Physics & Meteorology,* [2] *School of Medical Science & Technology*

*Indian institute of Technology., Kharagpur 721302*

[3] *WINGARD, Institute of Visual and Auditory Research, Kolkata 700012 ,India*

ABSTRACT

Gyrosonics refers to novel audio binaural stimulus that produces rotational perceptions of sound movement in head at a particular predetermined frequency. Therapeutic effect observed with this is considered to be associated with modification of arousal of autonomic nervous system. The heart rate variability (HRV), non-invasive measure of autonomic nervous system, has been measured for group of 30 subjects for pre- and post –gyrosonic installation. The time - and frequency- domain analysis of HRV results show overall decrease in sympathetic response and increase in para- sympathetic response due to listening of gyrosonics.

**Introduction**: Autonomous nervous system (ANS), with its two main divisions - sympathetic and parasympathetic may be viewed as hierarchically coordinated neuronal network in brain and plastically and rhythmically interfaced with brain, internal and external environments to regulate the exchange of energy and information [1]. The working of many internal organs including the heart, digestive tract, lungs, bladder and blood vessels is in turn regulated by the ANS. The sympathetic and the parasympathetic branches exert opposing effects on most organs. The sympathetic nervous system is rapidly activated in physically or mentally stressful situations and increases in HR, cardiac output, blood flow to the muscles, pupil dilation and a decrease in digestive system activity. It is therefore sometimes referred to as the 'fight or flight' response. On the other hand the parasympathetic induces opposing effects - HR and blood pressure (BP) to drop, the pupils to constrict and can be thought of as the 'rest and digest' mechanism [2]. Many of the disease entities (e.g. acute coronary syndrome, chronic heart failure, and diabetes mellitus) are characterized by pronounced sympathetic and parasympathetic imbalance [3]. It has long been known that harmonic motion can have an abundance of psychological and physical effects [4]. An auditory stimulus triggers a cascading series of events along the traveled route of stimulus response [5]. It is along this route where many of the beneficial effects of audio stimulus have their origins. As music modifies the psychobiological state of humans, it is believed that this connection can be utilized to have an impact on relieving stress related ailments [6].Gyro sonics–refers to audio stimulus with movement characteristic such that when applied binaurally it produces a perception of rotation of sound. The frequency of rotation is in infrasonic region. The perception of movement in auditory space by humans depends on a number of cues. It has been demonstrated that moving sound resulting from sequential excitation of the mono source through several speakers in free field can generate a specific activation in brain [7]. Sound motion evokes magnetic field and MEG study indicates that the right parietal cortex is involved in sound motion processing [8]. As gyro Sonics waves has much better rotational features compared to the experimentation done earlier [7, 8], it is expected that this can produce much larger spectrum of brain activation. Preliminary MEG studies with gyro Sonics have demonstrated that there is a motion-related magnetic response evoked by moving sound but not stationary sound [9]. Earlier study with this moving sound showed that the arousal level of psychosomatic patients were significantly reduced [10]. We have tried to find out the effect of gyro sonic sound on autonomic system and its possible therapeutic effects. Heart rate variability (HRV) measurements are normally used as markers of autonomous influence of heart. HRV can be assessed in time domain and in frequency domain. Time domain analysis provides quantitative measure of variation of heart rate, standard deviation, pNN50% (percent of difference between adjacent normal RR intervals that are greater than 50ms). Frequency domain analysis of HRV enables us to calculate the high frequency and low frequency power spectrum of fluctuation RR interval. It has been suggested that the low frequency power represents predominantly sympathetic modulation [11] and high frequency part is modulated by vagal activity [12]. We have chosen heart rate variability index to quantify the modulating influence of gyro Sonics on ANS.

**Materials and methods: Sample**: 30 volunteers from an information technology industry participated in this experiment. Physical and mental examination is within normal limit. All subjects' were right handed with no hearing disorder. All subjects gave informed written consent. The diurnal variation of HRV is also taken into consideration.

**Stimuli**: Stimuli for the experiment were pre-recorded gyrosonics constituted with rhythmic sound from percussion instrument of 7/8 (Indian Tabla). Stimuli were recorded digitally at a sample rate of 44.1 kHz and in 16 bit. In postproduction the amplitude of sound is modulated at 2 Hz (the proportion by which the amplitude was increased in one ear and decreased in the other) was created. Thus, approaching sound sources produce increases in intensity, and receding sound sources produce corresponding decreases, as occurs when sound moves in the horizontal plane around the head. The phase excursion was in advance at one ear and retardation at the other in .568-s condition. The slopes of the rising and falling sound amplitudes were exponential, and the stimuli were presented through headphones. Headphones were connected to a computer through a custom electronic interface. The sound was played at a sensation level of 50 db. The percept produced is of rotatory movement within a horizontal plane in head. The sound epoch was 9.5 min long. This duration was chosen to ensure that subjects clearly perceived the rotational movement of the sound. No subject experienced vertigo during or post playback condition. Experiments were carried out with the subjects in recumbent position in a soundproof room. All subjects underwent experiments in quite environment.

**Procedure:** BP and the 5 min ECG was taken for each Subject in each sitting and then they were exposed binaurally to rotating sound, following that again BP and 5-min ECG were taken. For ECG, Powerlab chart version 5.2.2 with ECG acquiring tool version 1.0.1 was used. The ECG data was of II lead and through 5 lead ECG acquiring systems. Ag/AgCl single use and latex free surface electrodes is used to get the signal. Following standard protocol [13] the heart rate variability was assessed. The different parameters defining HRV was obtained using RR-interval and computer programme. In time domain the heart rate (HR), root-mean squared successive differences (rMSSD) in NN (Normal to Normal)-interval and percentage of NN interval with a cycle length over 50ms different from the previous interval (pNN5) are obtained. All of these measurements of short-term variation estimate high frequency variations in heart rate and thus are highly correlated. In frequency domain power spectral density for low frequency (0.04 – 0.15 Hz) and high frequency (0.015 – 0.3Hz) are calculated. The distribution of power and the central frequency of LF and HF are not fixed but may vary in relation to changes in autonomic modulations of heart period.

**Results and Discussions:** The mean heart rate HR beats/min before (Pre) and after (Post) installation of audio stimulus is show in Fig.1.It is found that slowing down rate is more when initial value of beat is higher and almost no change is induced for lower HR. The mean HR of 30 persons changed from 82.2 b/m to 81.1 b/m. The systolic BP in mmHg before and after stimulus is presented. In most cases the response as regards BP is similar to that of HR. Subjects with Pre-BP within normal clinical limit had similar Post- BP.

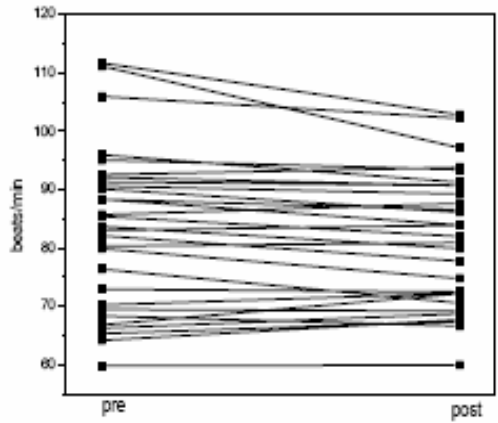

Fig1. Heart rate (b/m) before (Pre) and after (Post) application of gyrosonics

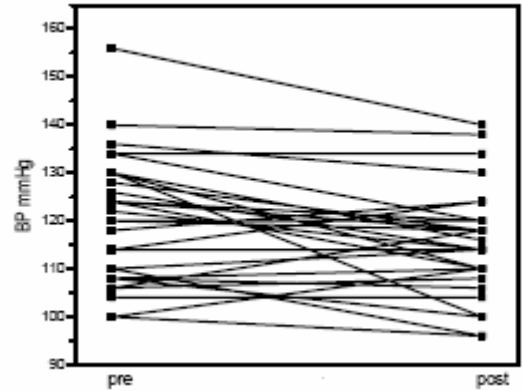

Fig.2 Systolic blood pressure BP mmHg before (Pre) and after (Post) application of gyrosonic

The mean systolic BP also reduced from 120 mmHg to 115 mmHg. This suggests that the gyro Sonics tends to facilitate restoration of the homeostatic condition.

Fig.3 and 4 depict the results of rMSSD (ms) and pNN50% in Pre- and Post- gyro Sonics. Majority response was such that rMSSD decreased after the audio listening. The mean value is changed from 36.4ms to 32.9ms. Similarly, pNN50 also altered significantly and mean decreases from 15.4 to 12.9 after gyro Sonics therapy. In contrast, changes in diastolic BP and standard deviation (SD) are non-significant.

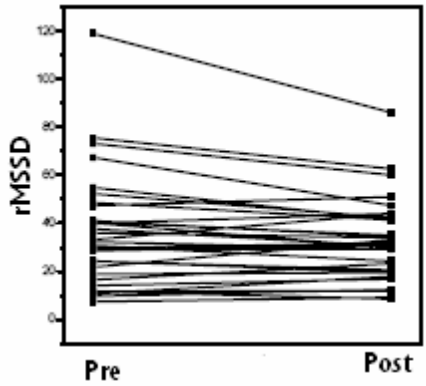

Fig.3 rMSSD (ms) before (Pre) and after(Post) application of gyrosonics

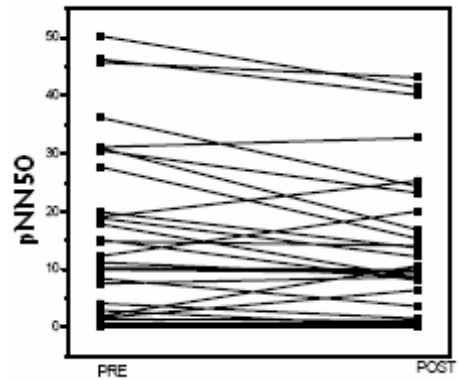

Fig.4. pNN50% before (Pre) and after(Post) application of gyrosonics

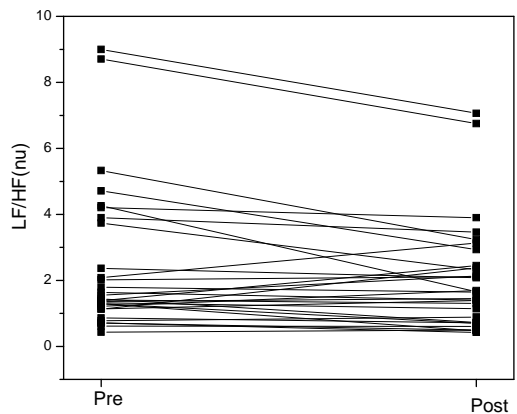

Fig.5 Ratio of LF/HF power in normalized unit at Pre- and Post gyrosonics.

Spectral power densities at LF and HF of NN interval are obtained for group at Pre- and Post- gyro Sonics. The LF component reflects both sympathetic and vagal input into heart whereas HF component appears to be originated from parasympathetic nervous system. A reciprocal balance is maintained such that when one branch of ANS is activated, the other one is in habited. For the purpose of study changes in power spectrum of HF are examined as the aim of the study is to ascertain the influence of gyrosonics on parasympathetic part of ANS. Fig. 5 displays ratio of LF/HF power in normalized unit for Pre- and Post session. There is a significant decrease in the ratio after therapy and indicates that gyrosonics can strongly enhance parasympathetic response. It is pertinent to note that similar study has been conducted with stationary (the rotational aspect has been excluded) same audio stimulus. The HRV indices exhibit non-significant changes before and after stationary audio session. Based on these observations it appears that the rotational part of sound is the key for modulation of autonomic control and related health benefit. It calls for further study with larger group to establish gyrosonics as novel non-invasive stimulant and the work is going on in this direction.

**Conclusions:** The results of the study show that there is significantly fall in HR, LF/HF ratio and BP of persons after gyrosonic therapy. It shows its potential implication on problems where there is autonomic imbalance e.g. hypertension, chronic heart failure, diabetes mellitus and anxiety or stress disorder. Fall in LF/HF shows that there is reduction in sympathetic component of ANS and effect of parasympathetic system gets augmented. Gyrosonics waves can be a good probe to study Brain and autonomic network. Further study on gyrosonic waves and its functional pathways to central autonomic oscillator through advanced Imaging modalities like fMRI and PET scan can further help us understanding the auditory motion perception.

Acknowledgement: The authors gratefully acknowledge  the Biomedical Signal analysis group, Department o Applied Physics ,Univ. of Kuopio,,Finland for allowing to use HRV software and express sincere thanks to  Dr. M.Oza,Dr.Amit and dr.S.Roy for assistance.